\documentclass[uplatex,12pt,a4paper]{jpsj3}%

\usepackage{overpic}
\usepackage{txfonts}
\usepackage{here}
\usepackage{comment}
\usepackage{tikz} 
\usepackage{graphicx}
\usepackage{dcolumn}
\usepackage{bm}
\usepackage{amsfonts}
\usepackage[mathlines]{lineno}
\usepackage{color}
\definecolor{Red}  {rgb}{1,0,0}
\definecolor{Green}{rgb}{0,1,0}
\definecolor{Blue} {rgb}{0,0,1}
\usepackage{bm}
\newcommand {\bfv}[1] {{\boldsymbol {#1}}}

 \usepackage{ulem}
 \usepackage{ifthen}

\title{Nonlinear Aggregation of Phase Elements on the Unit Circle under Parametric External Fields}

\author{Isshin Arai$^1$\thanks{k078403@kansai-u.ac.jp}, Tomoaki Itano$^2$, and Masako Sugihara-Seki$^2$}

\inst{$^1$Graduate School of Science and Engineering, Kansai University, Osaka, 564-8680, Japan\\
$^2$Department of Pure and Applied Physics, Faculty of Engineering Science, Kansai University, Osaka, 564-8680, Japan}

\abst{Inspired by flow visualizations by reflective flakes drifting in two-dimensional wave flows, we investigate nonlinear aggregation dynamics of phase elements distributed on the unit circle under a parametrically modulated external field.
Through numerical simulations and analytical approaches, we discover that the phase element model describing flake orientation motion, ${d\alpha/dt} = \lambda(t)\sin 2(\alpha - \phi(t))$ with $\lambda(t) = \cos(\omega_1 t)$ and $\phi(t) = \omega_2 t$, exhibits Arnold tongue-like structures in parameter space, where initially isotropic phase distributions aggregate into highly anisotropic states.
Complete aggregation occurs within contiguous wedge-shaped regions originating from a saddle-node bifurcation point.
The dynamics exhibit rich nonlinear behavior including attractors, limit cycles, and quasi-periodic trajectories in reduced indicator space spanned by aggregation degree ($I$), field-alignment measure ($O$), and temporal variation ($P$). Our findings reveal fundamental principles governing collective phase dynamics under competing temporal modulations, with potential applications spanning from biological synchronization to socio-economic dynamics and controllable collective systems.
}

\begin{document}
\maketitle

\section{Introduction}
  Flow visualization techniques using reflective flakes suspended in fluids constitute a common experimental method in fluid mechanics\cite{Dy82}.
  Flakes orient according to flow fields, and visualization patterns are generated through reflection, with these patterns critically depending on whether the local flake orientation distribution remains isotropic or becomes anisotropic\cite{Sa85, Gau98, Got11,Kida14}.
  While anisotropic distribution achieved locally may enable measurement of flow structures and characteristics, a primary question remains unresolved, i.e., under what flow conditions do flakes aggregate into anisotropic states rather than remain isotropically distributed?
  The motion of flakes originates from temporal changes in flow shear stress and vorticity along their Lagrangian trajectories, accompanied by difficulties in interpreting the conditions for flakes to aggregate into anisotropic states, which are also related to temporal variations of fluid elements \cite{Bat67,Gi90,Dr91}.
  
  When constrained to two-dimensional irrotational flows, the flake orientation dynamics can be mathematically reduced to phase elements motion on the unit circle.
  This reduction transforms the fluid mechanical problem into a dynamical systems framework, where each orientation angle of flake evolves according to the local velocity gradient field.
  This mathematical abstraction may connect flake aggregation dynamics observed in flow visualization experiments to a broader class of phase dynamical systems exemplified below.
  The connection between flake alignment in flows and the well-established synchronization phenomena\cite{Ad46, Ar65, Pi01, Guck13} has not yet been recognized.

Models describing synchronizing phase elements represented on the unit circle under external fields have been established as a framework in various fields. These models have been successfully applied to phenomena such as neural synchronization\cite{Br10}, chemical oscillators\cite{Ku03c}, and liquid crystal ordering\cite{Doi81, Hess76}.
A representative example is the Kuramoto model\cite{Ku1975}, which describes phase synchronization through inter-element interactions. Several physical systems involve phase elements with or without mutual interactions driven even by external fields, representing a distinct class of dynamics.
Moreover, external field driven models without inter-element interactions represent another class of dynamics, where characteristic motion patterns and spatial aggregation emerge from the temporal structure of external fields alone\cite{Je83}.
Such systems exhibit synchronization regions in parameter space that form structures reminiscent of Arnold tongues\cite{Pi01} and  a devil's staircase\cite{Je83}, related to nonlinear resonance and locking phenomena.

In the present study, we bridge fluid mechanics and nonlinear dynamics by revealing a relevance of flake orientation aggregation to resonance in dynamical systems.
We theoretically and numerically analyze the dynamical structure of phase element systems under temporally periodic-modulated external fields, demonstrating how established dynamical systems theory may provide quantitative predictions for fluid visualization outcomes while offering new physical realizations of locking phenomena.

In Section 2, we first formulate the governing equation of the rotation of flakes in two-dimensional irrotational flow.
In Section 3, we clarify conditions for multi-element distributions to converge to anisotropic states by numerical calculations.
Section 4 discusses the results through theoretical analysis, extended indicator space analysis, and extension to rotational flows.
Section 5 summarizes our findings and their broader implications.

\section{Formulation and Methodology}
To address the fundamental question of aggregation conditions posed in the introduction, we examine a mathematically tractable case of two-dimensional irrotational flows. Despite this simplification, the system exhibits unexpectedly complex parameter dependencies that warrant detailed investigation.

Following Refs.~\cite{Je1922,Bat67,Got11} the three-dimensional orientation of infinitesimal particles in a three-dimensional flow field can be governed by the Jeffery equation. In the limit of extremely flat particles (zero thickness), the equation reduces to $\dot{\bfv{s}} = \bfv{s} \times (\bfv{s} \times (\bfv{\nabla}\bfv{u} \cdot \bfv{s}))$, where $\bfv{s}$ is the unit normal vector of the flake and $\bfv{\nabla}\bfv{u}$ is the velocity gradient tensor at the flake position.
In a two-dimensional irrotational flow (potential flow), the velocity gradient tensor is symmetric with orthogonal principal  axes of real eigenvalues, and trace zero due to incompressibility.
One principal direction at a time $t$ is represented by the argument $\phi(t)$ on the flow plane.
The symmetric velocity gradient tensor $\nabla \bfv{u}$ in three-dimensional space has an eigenvalue $\lambda(t)$ with corresponding normalized eigenvector $\bfv{e}_{\lambda} \equiv \cos\phi(t)\, \bfv{e}_x + \sin\phi(t)\, \bfv{e}_y$,
  where $\bfv{e}_x$ and $\bfv{e}_y$ are orthogonal unit vectors on the flow plane.
  Then, the velocity gradient tensor $\nabla\bfv{u}$ can be calculated from the following matrix product,
\begin{align*}
  \begin{pmatrix}
    \cos\phi(t) & -\sin\phi(t) \\
    \sin\phi(t) & \cos\phi(t)
  \end{pmatrix}
  \begin{pmatrix}
    \lambda(t) & 0 \\
    0 & -\lambda(t)
  \end{pmatrix}
  \begin{pmatrix}
    \cos\phi(t) & \sin\phi(t) \\
    -\sin\phi(t) & \cos\phi(t)
  \end{pmatrix}.
\end{align*}

Let $\alpha(t)$ denote the argument of the vector $\bfv{s}$ on the two-dimensional flow plane, defined by $\bfv{s} \equiv \cos{\alpha}\, \bfv{e}_x + \sin{\alpha}\, \bfv{e}_y$.
Substituting the expression of $\bfv{\nabla}\bfv{u}$ by $\phi(t)$ and $\lambda(t)$ 
into the aforementioned orientation equation leads to
the evolution equation of $\alpha$ describing a single flake orientation.
Whether the flakes with different initial conditions asymptotically align in the same direction or remain dispersed under the common velocity field directly affects the reflection intensity observed in flow visualization patterns.
We now assume a collection of $N$ flakes within a sufficiently small fluid region such that all flakes experience the same local velocity gradient field.
While each flake has its own initial orientation, they should be all governed by the same underlying flow dynamics.
Therefore, the governing equation of the argument $\alpha_i(t)$ of the $i$-th flake ($i = 1, 2, \cdots, N$) is 
\begin{equation}
\frac{d{\alpha}_i}{dt} = \lambda(t) \sin 2(\alpha_i - \phi(t)) \ .
\label{eq:model}
\end{equation}
The invariance of this equations under the transformation $\alpha \to \alpha + \pi$ is important, which is related to the equivalence between the front and back sides of a flake.

As a motivating physical system, let us imagine water waves.
Linearized water wave theory predicts that fluid elements near the surface follow elliptical trajectories with an angular frequency $\tilde{\omega}$ when the wavelength is sufficiently short compared to the depth.
In this case, flakes suspended in the fluid experience periodic reorientation under a time-varying velocity gradient.
Deeper in the fluid, however, horizontal motion dominates, and the strength of the velocity gradient varies periodically.
While flake orientation dynamics near the water surface involve constant rotational strength, those close to the bottom are expected to experience time-varying principal direction with a modulation.
Specifically, near the surface, the eigenvalue magnitude remains approximately constant, where the principal directions rotate with a constant angular frequency.
On the other hand, near the bottom, both of them exhibit highly distorted temporal oscillations: the eigenvalue shows a periodic variation with the double frequency $2\tilde{\omega}$ while maintaining its constant sign, and the principal direction undergoes periodic modulation with the frequency $\tilde{\omega}$.
This physical foundation motivates us to introduce two independent frequencies in our model, characterizing the eigenvalue oscillation and the principal direction rotation.
While these parameters are related in actual water waves, we treat them as independent to explore the full parameter space of the underlying dynamical system.

Building upon insights from the water wave analogy, we adopt a velocity gradient field with two independent periods:
\begin{equation}
  \lambda(t) = \cos(\omega_1 t), \quad \phi(t) = \omega_2 t.
  \label{eq:model-case}
\end{equation}
In the context of phase elements synchronization phenomena, we can reinterpret this equation on the flake orientation as a phase dynamical system interacting with a "switching rotating attractive device" - an external field where the {\it attractive point} rotates at constant angular velocity while the {\it attraction strength} oscillates between positive and negative values.
The values of $\omega_1$ and $\omega_2$ represent the inverse of modulation time of signed attractive strength and the rotation period of the attractive point, respectively. While water wave dynamics would typically impose rational frequency relationships, we hereafter treat $\omega_1$ and $\omega_2$ as independent parameters to explore the full dynamical structure.
When time variation in attractive strength and attractive point are sufficiently slow, it is intuitively understood that phase elements aggregate toward the attractive point.
However, as their time variations increase, such aggregation behavior becomes non-trivial, with possibilities of complex aggregation phenomena.

\begin{figure}
    \centering
    \includegraphics[width=0.8\linewidth]{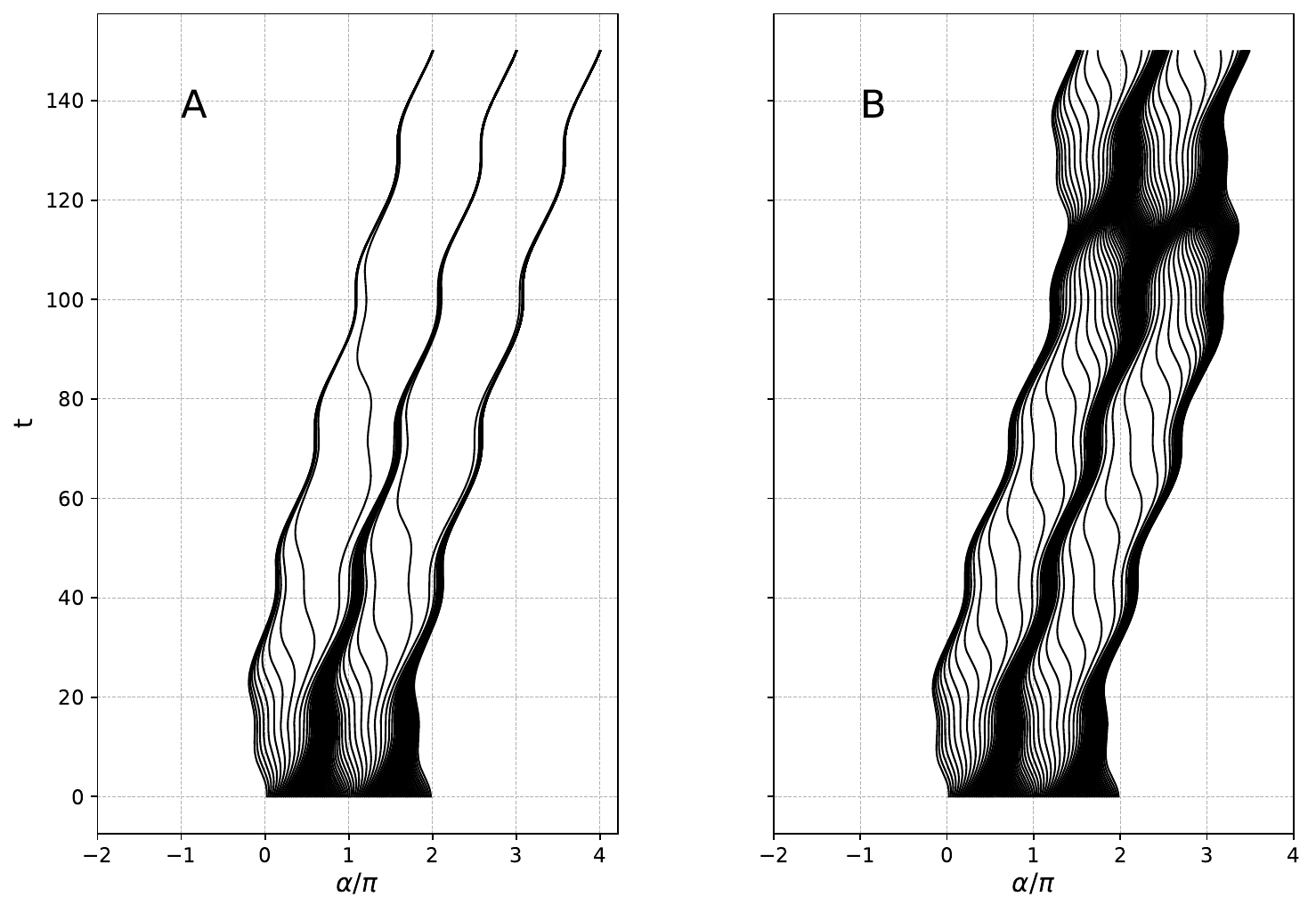}
    \caption{
      Time evolution of $N=100$ phase elements initially distributed isotropically at $t=0$. Left panel (Case A) shows $(\omega_1, \omega_2) = (1.1, 1.1)$ where phases $\alpha$ converge to an anisotropic state over time. Right panel (Case B) shows $(\omega_1, \omega_2) = (1.1, 1.2)$ where the system periodically alternates between isotropic and anisotropic states.
    }
    \label{fig:fig-0}
\end{figure}

Figure~\ref{fig:fig-0} demonstrates the time evolution of $\alpha_i(t)$ for $i = 1, 2, \cdots, N$ ($N=100$), where the phase elements are initially distributed isotropically, at the case A: $(\omega_1, \omega_2) = (1.1, 1.1)$ and the case B: $(\omega_1, \omega_2) = (1.1, 1.2)$.
For the case A, phase elements gather over time and eventually form an anisotropic state (Fig.~\ref{fig:fig-0}A).
In contrast, for the case B, they gather once but then return to isotropy (Fig.~\ref{fig:fig-0}B).
As demonstrated by this contrast, our primary interest lies in whether multiple phase elements moving on the unit circle according to Eq.(\ref{eq:model}) under the condition (\ref{eq:model-case}) can experience a convergence from an initially isotropic distribution to an aggregated, anisotropic state as time evolves.
Thus, we introduce the aggregation degree $I \in [0,1]$:
\begin{equation}
I = \text{tr}(\mathcal{D})^2 - 4\det(\mathcal{D}),
\label{eq:eq-I}
\end{equation}
where the orientation tensor $\mathcal{D}$ is calculated from
\[
  \frac{1}{N} \sum_{i=1}^N \begin{pmatrix} \cos^2\alpha_i & \cos\alpha_i\sin\alpha_i \\ \cos\alpha_i\sin\alpha_i & \sin^2\alpha_i \end{pmatrix}.
\]
This plays a similar role as order parameters in Kuramoto theory and Doi-Hess theory. The invariance of Eq.(\ref{eq:model}) under the transformation $\alpha \to \alpha + \pi$ also reflects the symmetry of the aggregation degree.
When the distribution at phase $\alpha$ is expressed as $f(\alpha)d\alpha$ using the probability density function $f(\alpha)$ over the interval $\alpha$ to $\alpha + d\alpha$, and in the limit $N \to \infty$, the value of $I$ in Eq.~(\ref{eq:eq-I}) is given by $I = \left( \int_0^{2\pi} \cos{2\alpha} \, f(\alpha) \, d\alpha \right)^2 + \left( \int_0^{2\pi} \sin{2\alpha} \, f(\alpha) \, d\alpha \right)^2$. Noting that the expression of $I$ corresponds to Fourier expansion coefficients, $I$ becomes 0 not only for isotropic distributions but also for any $f(\alpha) = \frac{1}{2\pi} + \sum_{n=2}^{\infty} a_n \cos(2n\alpha) + b_n \sin(2n\alpha)$. Furthermore, under $\int_0^{2\pi} f(\alpha) d\alpha = 1$ and $f(\alpha) > 0$, $I$ takes its maximum value of 1 when all phase elements $\alpha_i$ coincide at a single angle $\alpha_0(t)$, i.e., $f(\alpha) = \delta(\alpha - \alpha_0)$. Thus, $I$ serves as an indicator quantifying the aggregation degree of the phase elements.

A commonly used complex order parameter for a single phase variable is defined as $\Psi = \left| \frac{1}{N} \sum_{j=1}^{N} e^{i \alpha_j} \right|$.
To reflect the invariance of Eq.~(\ref{eq:model}) under \(\alpha_j \rightarrow \alpha_j + \pi\), it can be modified to $\Psi' = \left| \frac{1}{N} \sum_{j=1}^{N} e^{i 2 \alpha_j} \right|$.
This order parameter can be interpreted geometrically as the distance from the origin to the centroid of the phase points on the unit circle.
In the present study, the aggregation degree \(I\) is essentially equivalent to \(\Psi'\).

\section{Results}

\begin{figure}
    \centering
    \includegraphics[width=0.8\linewidth]{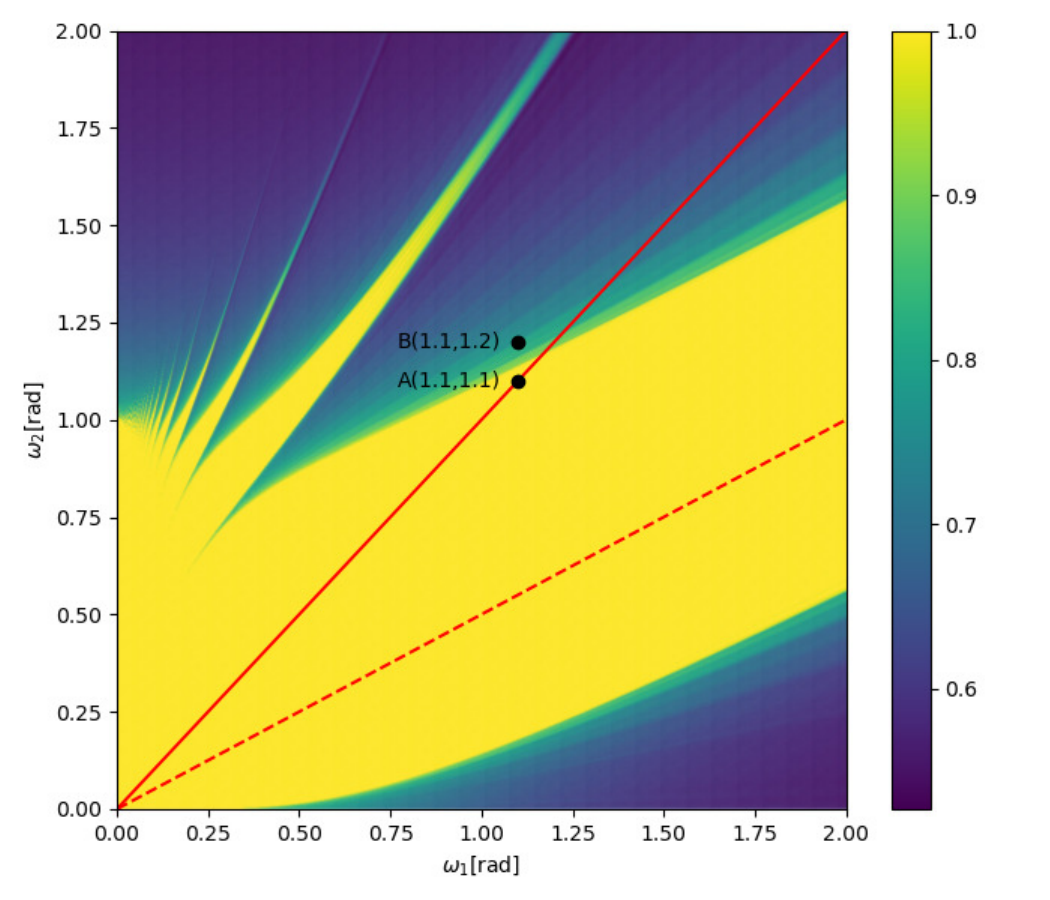}
    \caption{
    Contour map of $\langle I \rangle_{[T/2, T]}$ for $N=20$.
      Horizontal and vertical axes are $\omega_1$ and $\omega_2$, respectively.
      Parameter resolution is $\Delta\omega = 2.0\times{10^{-3}}$.
      Solid and dashed red lines indicate $\omega_1 = \omega_2$ and $\omega_1 = 2\omega_2$, respectively.
      In the bright region where $\langle I \rangle_{[T/2, T]} \approx 1$, all phase elements completely aggregate, forming contiguous wedge-shaped regions that originate from (0,1), with a prominent band having slope $1/2$.
      The red solid line $\omega_1 = \omega_2$ intersects the boundary of the bright region at $\omega_1\approx 1.16$.
      Points A and B correspond to Fig.~\ref{fig:fig-0}A and B, respectively.}
    \label{fig:fig-2}
\end{figure}

To investigate this aggregation behavior systematically, numerical simulations of Eq.(\ref{eq:model}) were performed using fourth-order Runge-Kutta method with a time step $\Delta t=10^{-2}$ over the time interval from $0$ to $T$ with $T=10\cdot2\pi/\omega_1$ for $\omega_1\ne 0$, otherwise $T=500$. The number of elements \(N\) varies depending on the specific simulation performed. The value of \(N\) used in each case is explicitly stated where relevant. Figure~\ref{fig:fig-2} shows contours of the time average $\langle{I}\rangle_{[T/2,T]}$ of $I(t)$ from $T/2$ to $T$ on the $\omega_1$-$\omega_2$ plane examined with the parameter resolution $\Delta\omega=2.0\times10^{-3}$ for $N=20$. 
Anisotropic state $\langle I \rangle \approx 1$ is realized in a bright region, while persistent isotropy $\langle I \rangle \approx 0$ remains over sufficient washout time out of the region.

From this contour plot, we first observe that anisotropic state of $N$ phase elements is attained  in a band with a finite width with slope $1/2$ in the parameter space.
The term 'aggregation' refers to the phenomenon where phase elements converge toward similar orientations, which may be either transient or persistent. When all phase elements, regardless of their initial orientations, converge to a single common periodic trajectory with constant phase offsets and never return to isotropy thereafter, as illustrated in Fig.~\ref{fig:fig-0}(A), this will be termed {\it complete aggregation}. 
As shown in Fig.~\ref{fig:fig-2}, complete aggregation occurs over a continuous band region, revealing nontrivial structure in the parameter plane.
In other words, the complete aggregation of phase elements occurs in such continuous parameter range, that is, the inverse function for obtaining parameters $\omega_1$ and $\omega_2$ from the aggregation indicator $I$ becomes multivalued, so that parameter estimation based on $I$ is impossible. Secondly, contiguous wedge-shaped regions with discrete slopes,  as reminiscent of a maple leaf prominently showing the nonlinearity of the entire system, are confirmed to originate from $(\omega_1,\omega_2)=(0,1)$.

\begin{figure}[htbp]
    \centering
    \begin{minipage}{0.8\linewidth}
        \centering
        \includegraphics[width=\textwidth]{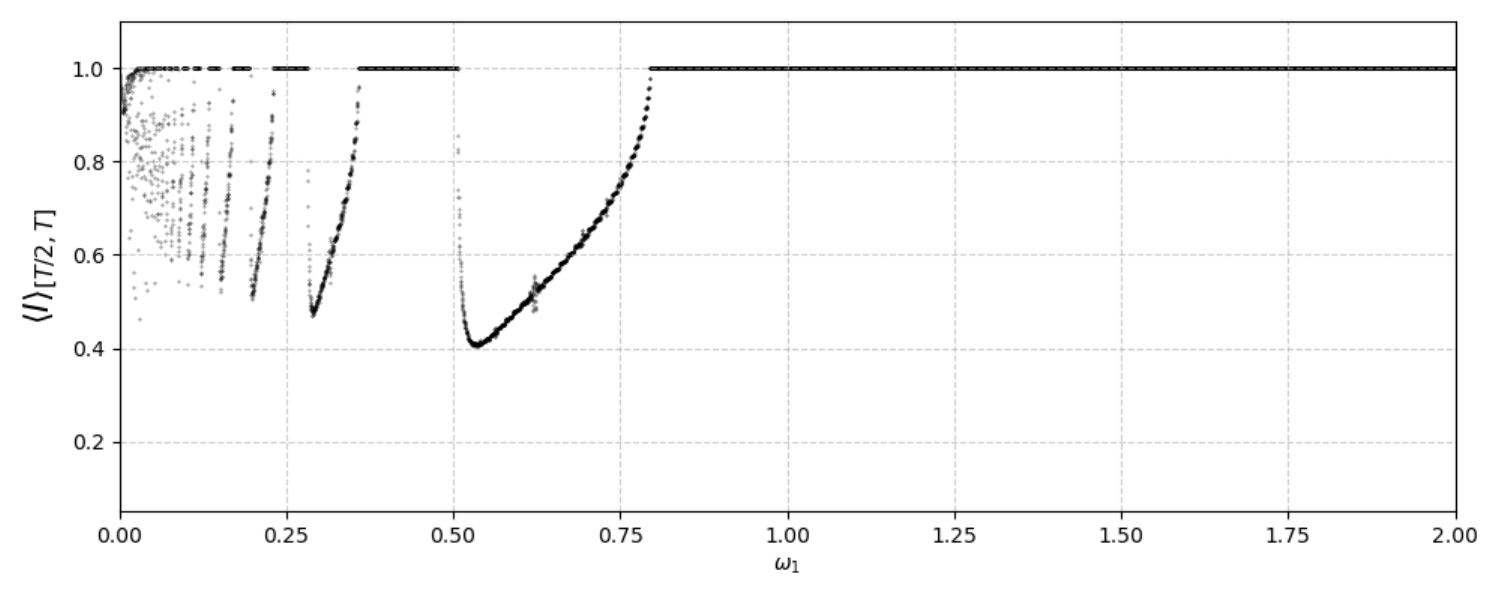}
        {\small (a): $\omega_2=1$} 
    \end{minipage}

    \vspace{1em} 

    \begin{minipage}{0.8\linewidth}
        \centering
        \includegraphics[width=\textwidth]{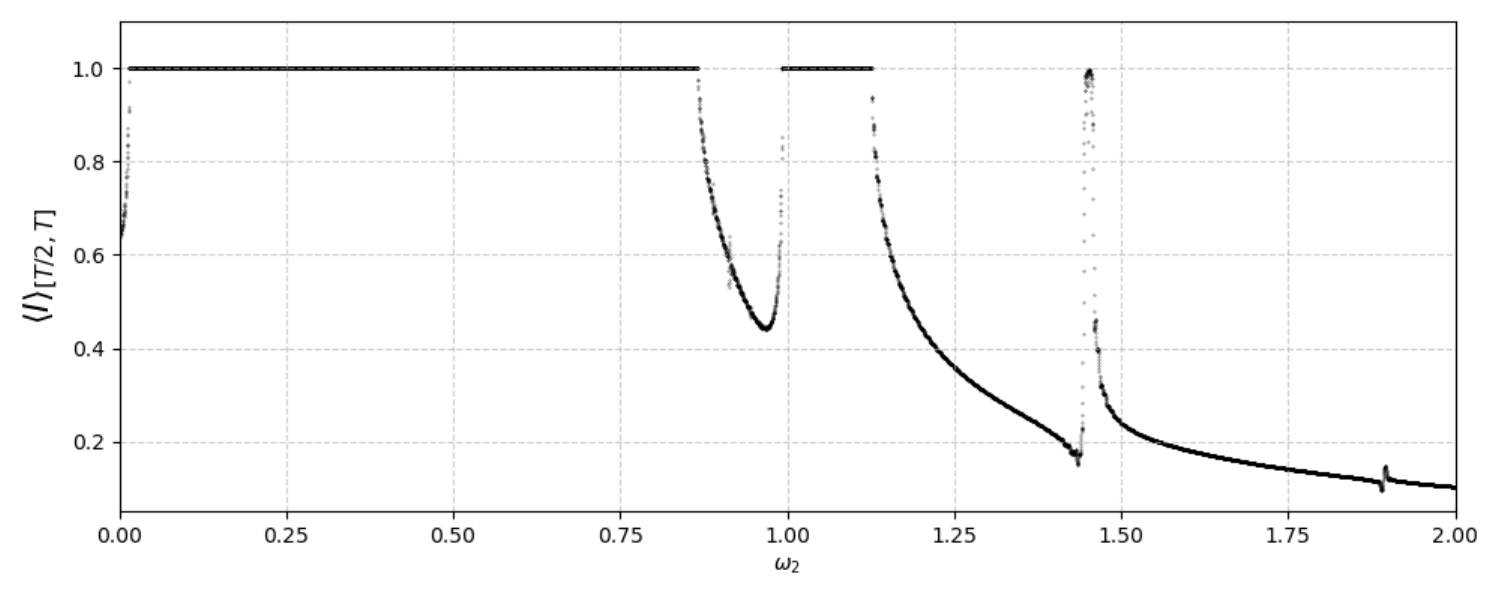}
        {\small (b): $\omega_1=1/2$} 
    \end{minipage}

    \caption{
Representative slices of $\langle I \rangle_{[T/2,T]}$ for $N=20$, illustrating the detailed behavior near the critical regions.

Case (a) ($\omega_2 = 1$): Multiple plateau regions with $\langle I \rangle \approx 1$ are observed, each followed by sharp departures near the corresponding critical points.

Case (b) ($\omega_1 = 1/2$): Similarly, multiple plateau regions are present, each followed by sharp departures near the corresponding critical points.

These slices were computed with slightly different parameters 
($\Delta\omega=2.0\times 10^{-4}$, $T=50 \cdot 2\pi / \omega_1$) for higher resolution.
}
    \label{fig:fig-slice}
\end{figure}

While the contour plot in Fig.~\ref{fig:fig-2} provides a global view of the aggregation behavior, it does not fully resolve the detailed bifurcation structure near the critical regions. To clarify this point, Fig.~\ref{fig:fig-slice} shows representative slices of $\langle I \rangle_{[T/2,T]}$ for $N=20$, illustrating the detailed behavior near the critical regions.
As shown in Fig.~\ref{fig:fig-slice}(a), when \(\omega_2 = 1\) is fixed, multiple plateau regions with \(\langle I \rangle \approx 1\) appear, each followed by sharp departures near the corresponding critical points.  
Similarly, in Fig.~\ref{fig:fig-slice}(b), when \(\omega_1 = 1/2\) is fixed, multiple plateau regions are observed, again followed by sharp departures near the critical points. From a qualitative evaluation of these slices, it appears that near the critical points, the order parameter $\langle I \rangle$ exhibits a power-law–like variation. The steepness of this change is observed to depend on both the location of the critical point and the control parameter being varied. Defining $\delta \langle I \rangle \equiv  1- \langle I \rangle$, this power-law–like behavior is reflected in the variation of $\delta \langle I \rangle$. These variations correspond to how strongly the observed patterns or light intensity change with the control parameters.

Also, at $\omega_1 \neq 0$ and $\omega_2 = 0$, the solutions are periodic, given by $\alpha = \tan^{-1}\!\left(\exp{({{2\sin(\omega_1 t)}/{\omega_1}})} \tan{\alpha(0)}\right)$.
Consequently, the complete aggregation region does not extend to the $\omega_1$-axis.

To relate the complete aggregation to the well-known Arnold tongues, we extend our model by introducing a nonlinearity positive parameter $\epsilon$,
  \begin{equation*}
    \frac{d\alpha}{dt} = \epsilon\cos(\omega_1 t)\sin(2(\alpha - \omega_2 t)).
  \end{equation*}
Note any system with $\epsilon \neq 1$ can be reduced to the standard case $\epsilon = 1$ by rescaling time with replacing parameters $(\omega_1,\omega_2) \to (\omega_1/\epsilon,\omega_2/\epsilon)$.
When $\omega_1 = 0$, via $\tilde{\alpha}=\alpha-\omega_2 t$, this equation reduces to Adler-type equation, $ \frac{d\tilde{\alpha}}{dt} = -\omega_2 + \epsilon\sin{2\tilde{\alpha}}$, which describes the synchronization of autonomous oscillating system to an external driving force. 
Additionally, a saddle-node bifurcation occurs at $\omega_2=\pm\epsilon$ in the $\tilde{\alpha}$ dynamics, marking the transition between fixed and periodic behaviors.
    
\begin{figure}[htbp]
    \centering
    \begin{minipage}{0.45\linewidth}
        \centering
        \includegraphics[width=\textwidth]{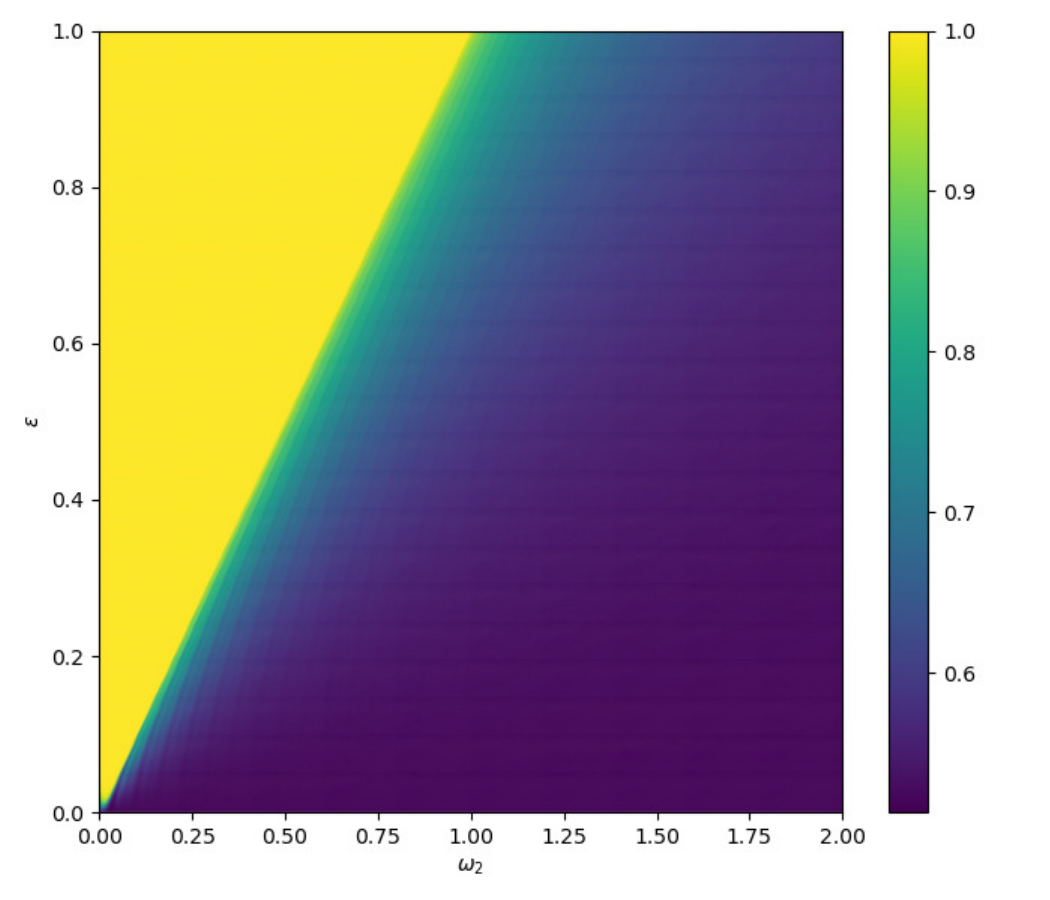}
        {\small (a):$\omega_1=0$}  
    \end{minipage}
    \hfill
    \begin{minipage}{0.45\linewidth}
        \centering
        \includegraphics[width=\textwidth]{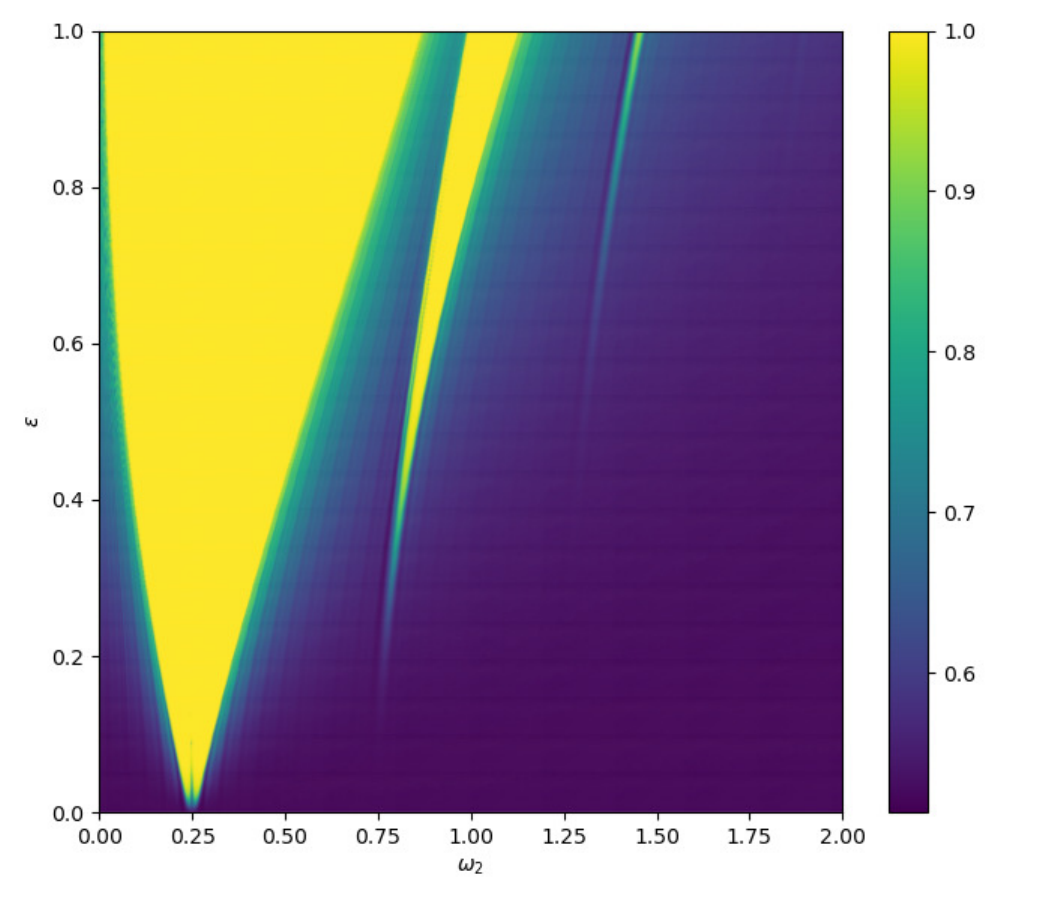}
        {\small (b):$\omega_1=1/2$} 
    \end{minipage}
    \caption{
Contour plot of
${\langle I \rangle_{[T/2,T]}}$ in the $\omega_2$–$\epsilon$ plane 
for $N=20$, illustrating how the aggregation regions depend on the $\omega_1$.

Case (a) ($\omega_1=0$): 
A single tongue-shaped region with 
${\langle I \rangle_{[T/2,T]} \approx 1}$ appears in the domain $\epsilon > \omega_2$. 

Case (b) ($\omega_1=1/2$): 
Multiple tongue-shaped regions with 
${\langle I \rangle_{[T/2,T]} \approx 1}$ emerge.
}
    \label{fig:fig-Arnoldtongue0}
\end{figure}

On the $\omega_2$-$\epsilon$ plane, the complete aggregation region for $N=20$ in the $\tilde{\alpha}$ dynamics is characterized as the classical Arnold tongue as shown Fig.~\ref{fig:fig-Arnoldtongue0}(a), exhibiting a structure similar to the synchronization region.
For $\omega_1=0$, the Adler-type equation\cite{Ad46} displays a single tongue, and contour analysis of the winding number $\rho=(\tilde{\alpha}(T)-\tilde{\alpha}(0))/T$ confirms the formation of identical patterns.
When $\omega_1$ is $1/2$, $\langle{I}\rangle_{[T/2,T]}$ on the $\omega_2$-$\epsilon$ plane in Fig.~\ref{fig:fig-Arnoldtongue0}(b) demonstrates the emergence of multiple tongue-shaped regions, which correspond to the wedge-shape  in Fig.~\ref{fig:fig-2}.

The contour in Fig.~\ref{fig:fig-2} can thus be understood as follows: when we examine $I$ values by varying $\epsilon$ (effectively zooming into different scales), the resulting $I$ pattern on the $\omega_2$-$\epsilon$ plane corresponds to the tip part of an Arnold tongue as shown in Figs.~\ref{fig:fig-Arnoldtongue0}(a) and (b).
Moving toward larger $\epsilon$ values, we approach the ``throat'' of the Arnold tongue.
Thus, the shape observed on the $\omega_1$-$\omega_2$ plane of Fig.~\ref{fig:fig-2} including the maple leaf-like pattern corresponds to the throat of the Arnold tongues, where multiple resonance conditions apparently interact.
This geometric interpretation explains why the wedge-shaped regions originate from the saddle-node bifurcation point.

\section{Discussion}

\subsection{Analytical Study of Phase Aggregation}
In the previous section, the aggregation phenomena were identified across a range of continuous parameter values, where the numerical results showed that complete aggregation occurs primarily at rational frequency ratios, for example, $\omega_1:\omega_2=2:1$ or $1:1$, which may appear naturally in wave phenomena.
Here, we develop a theoretical analysis to explain how aggregation emerges at these specific parameter values, focusing on these two representative cases.

\subsubsection{Case 1($\omega_1 = 2\omega_2$)}
Within the band region containing the dashed line with slope 1/2 (Fig.~\ref{fig:fig-2}) introduced in Section 2, we observed the complete aggregation.
In this region, each phase element $\alpha$ oscillated periodically around a slowly drifting center $\alpha_{\rm c}(t)$.
The drift rate of $\alpha_{\rm c}(t)$ decreases as the parameters approach the special line $\omega_1=2\omega_2$, where $\alpha_{\rm c}(t)$ becomes completely stationary.
Furthermore, the amplitude of the oscillation was found to gradually decrease as $\omega$ increased.

We analyze the dynamics of the phases in the case where $\omega_1 = 2\omega_2 = \omega$.
Under this condition, the Eqs. (\ref{eq:model}) and (\ref{eq:model-case})  reduce to
\begin{equation*}
\frac{d\alpha}{dt} = \cos(\omega t) \sin\left(2\alpha - \omega t\right).\label{eq:omega1_omega2}
\end{equation*}
To account for the residual oscillations of \(\alpha(t)\) in the high-frequency limit, we decompose the phase variable into a slowly varying component \(\alpha_\text{c}(t)\) and a small, rapidly oscillating correction \(\varphi(t)\), such that
\[
\alpha(t) = \alpha_\text{c}(t) + \varphi(t), \quad \text{with} \quad |\varphi(t)| \ll 1.
\]
Here, \(\alpha_\text{c}(t)\) captures the long-term dynamics, while \(\varphi(t)\) represents fast oscillations with approximately zero mean over one period of the fast time scale.
Introducing the fast time variable \(\tau = \omega t\), the equation can be approximated as
\[
\frac{d\alpha_\text{c}}{dt} + \frac{d\varphi}{dt} = \cos(\tau)\left[\sin(2\alpha_\text{c} - \tau) + 2\varphi \cos(2\alpha_\text{c} - \tau)\right].
\]
We now apply averaging over the fast time \(\tau\) to extract the effective dynamics of \(\alpha_\text{c}(t)\). Since \(\varphi\) was assumed to have zero average and the product \(\varphi \cos(\tau)\) yields only oscillatory contributions, the second term on the right-hand side vanishes upon averaging. Thus, we obtain the effective equation:
\[
\frac{d\alpha_\text{c}}{dt} = \frac{1}{2\pi} \int_0^{2\pi} \cos(\tau) \sin(2\alpha_\text{c} - \tau)\, d\tau = \frac{1}{2} \sin(2\alpha_\text{c}).
\]
The fixed points of the averaged equation satisfy \(\sin(2\alpha_\text{c}) = 0\), which implies \(\alpha_\text{c} = n\pi/2\) for \(n \in \mathbb{Z}\). Among these fixed points, those given by \(\alpha_\text{c} = \pi/2 + m\pi\), with \(m \in \mathbb{Z}\), are stable equilibria. 
In the limit \(\omega \to \infty\), the fast oscillatory correction vanishes, i.e., \(\varphi(t) = 0\), and thus \(\alpha(t)\) coincides with the fixed point \(\alpha_\text{c}\).
This matches the previous numerical observations.

\subsubsection{Case 2($\omega_1 = \omega_2$)}
Having examined the $2:1$ resonance, we now analyze the equal-frequency case $\omega_1=\omega_2=\omega$.
Among the various frequency combinations possible in linear wave superposition, this ratio constitutes another fundamental resonance condition and offers a complementary perspective to the previous analysis.
\begin{equation*}
\frac{d\alpha}{dt} = \cos(\omega t) \sin2\left(\alpha - \omega t\right). 
\end{equation*}

Unlike the $2:1$ case where high-frequency averaging was effective, the equal-frequency condition requires a different approach.
From Fig.~\ref{fig:fig-2}, we expect that as $\omega$ increases, the system will eventually leave the complete aggregation solid line, resulting in a loss of aggregation.
To analyze this transition, we introduce sum and difference variable, $x = \alpha_1 - \alpha_2, \quad y = \alpha_1 + \alpha_2 $.
The equations of motion transform to the following three-dimensional nonlinear ordinary differential equations.
\[
\Bigl(\frac{dx}{dt},\frac{dy}{dt},\frac{dz}{dt}\Bigr) = \Bigl( 2\cos z \sin x \cos(y - 2z),  2\cos z \cos x \sin(y - 2z), \omega \Bigr).
\]
Complete aggregation occurs when the difference coordinate $x$ remains stable near $x=0$,  ensuring convergence to $\alpha(t)=y(t)/2$.

Numerical simulations for this system revealed a bifurcation associated with the loss of stability of the invariant set $x=0$ at the critical value $\omega = \omega_{\rm cr} \approx 1.164$.  
In the segment $\omega \leq \omega_{\rm cr}$, the equilibrium $x = 0$ is stable, ensuring that initially different $\alpha_1$ and $\alpha_2$ converge to the same value. The resulting $\alpha(t) = y(t)/2$ exhibits periodic behavior with period $\pi/\omega$.

\ \\

With a parameter $k$ defined by $\omega_2 = k \omega_1$, the case $k=1/2$ (Section 4.1.1) showed no bifurcation, $k=1$ (Section 4.1.2) revealed a bifurcation point at a finite $\omega_{\rm cr}$, located on the boundary of the contiguous wedge-shaped regions in the parameter space (Fig.~\ref{fig:fig-2}).
This approach suggests a unified parametric study by varying $k$.
As $k$ increases systematically, the critical points $(\omega_1,\omega_2) = (\omega_{\rm cr}(k), \omega_{\rm cr}(k)/k)$ obtained within this framework are expected to trace a curve in the parameter plane. This parametric curve would naturally connect to the critical condition of the Adler-type equation in the limit $k \to \infty$.

In summary, the analyses of both cases of different resonance, $\omega_1:\omega_2 = 2:1$ and $1:1$, show that complete aggregation states can emerge under different frequency relations. These analytical results complement the numerical findings by explaining the emergence of aggregation within the band-shaped parameter regions, thereby confirming that the observed structures arise from the underlying nonlinear dynamics rather than from numerical artifacts. Moreover, by reducing the system to Adler-type equation, the existence of the band can be further supported by the classical synchronization theory\cite{Ad46,Bh09}.

\subsection{Extended Analysis in IOP Space}
In this section,
we examine the system response against the external field by introducing two additional measures.
In addition to aggregation degree $I$, we denote a measure of alignment between $\alpha$ and $\phi(t)$ as $O$,
and its time variation as $P$.
\[
O = \frac{1}{N} \sum_i |\cos(\alpha_i - \phi(t))| \ \ \ , \ \ \ 
P = \frac{dO}{dt},
\]
where $N$ is the total number of phase elements.
The value of $O$ measures the alignment between phase elements and the external field direction $\phi(t)$.
Since eigenvector directions are defined up to sign (both $\bfv{e}_\lambda$ and $-\bfv{e}_\lambda$ are equivalent), we adopt absolute values to account for this directional ambiguity.
This quantity ranges from 0 (perfect anti-alignment) to 1 (perfect alignment),
providing a direct measure of how well the phase distribution follows the rotating attractive point.
The space $\mathbb{A}$ spanned by $(I, O, P)$ provides another reduced representation of the multi-element dynamics, in which the trajectories capture the system's response to different external field conditions as well as to their initial conditions.

Space $\mathbb{A}$ defined in this way should be interpreted not as a static feature space but as a reduced dynamical structure space, reflecting only the limited information available through $(I,O,P)$. 
While some details of the collective distribution are inevitably lost, trajectories in $\mathbb{A}$ still capture history-dependent responses to external fields. Under specific conditions, they may form attractor-like structures. In general, their evolution depends on the initial conditions.

\begin{figure}
    \centering
     \includegraphics[width=0.8\linewidth]{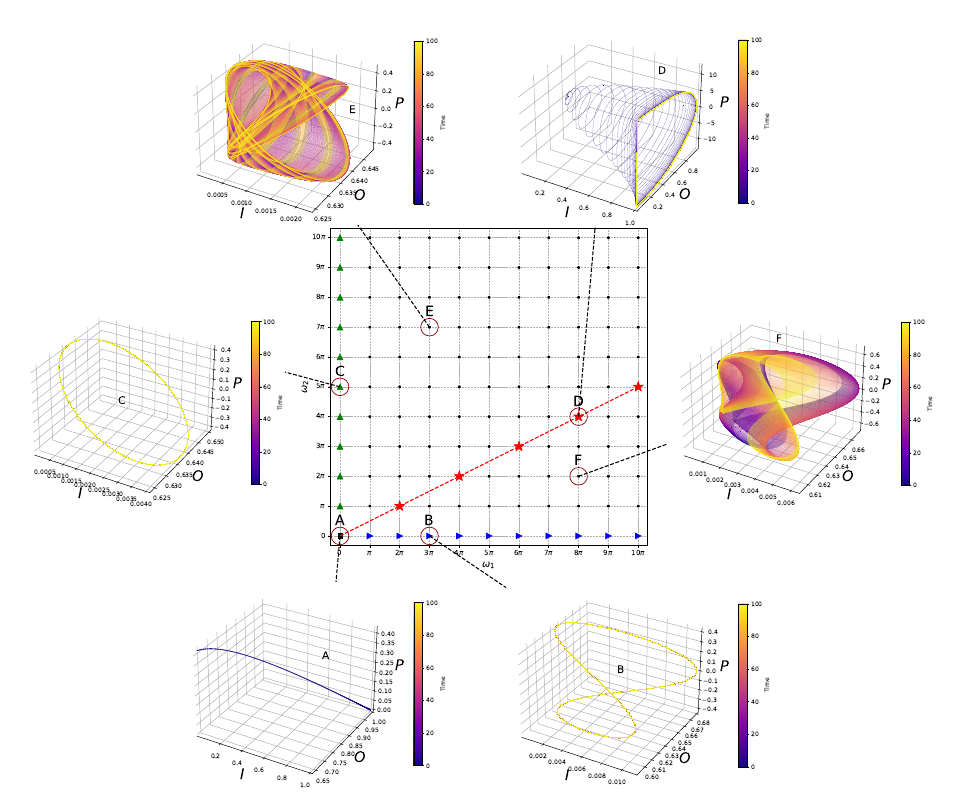}
    \caption{Examples of $(I,O,P)$ trajectories for various $\omega_1$, $\omega_2$ settings, starting from isotropic initial orientations of $\alpha_i$ for $N=5$,$000$.
(A) $\omega_1 = \omega_2 = 0$: Convergence to attractor $(1,1,0)$ under static external field.
(B) $\omega_1 \neq 0$, $\omega_2 = 0$: Periodic motion with $O$-$P$ describing closed orbit.
(C) $\omega_1 = 0$, $\omega_2 \neq 0$: Elliptical periodic orbit.
(D) $\omega_2/\omega_1 = 1/2$: Limit cycle in complete aggregation state.
(E,F) Complex quasi-periodic trajectories by superposition of basic orbits.}
\label{fig:fig-phase}
\end{figure}

Investigation of time evolution for grid points $n\pi$ $(n = 0, 1, \dots, 10)$ on parameters $(\omega_1, \omega_2)$, starting from isotropic initial orientations of $\alpha_i$ for $N=5$,$000$, confirmed several geometric features (Fig.~\ref{fig:fig-phase}).
First, for $\omega_1 = \omega_2 = 0$, i.e., $\alpha = \tan^{-1}(e^{2t} \tan{\alpha(0)})$, $(I,O,P)$ converges to an attractor $(I, O, P) = (1, 1, 0)$. In this case, external fields experienced by collective phase elements do not change temporally, and the relationship between external field and aggregation point remains constant (Fig.~\ref{fig:fig-phase}A).
Next, when $\omega_1 \neq 0$, $\omega_2 = 0$, we have $\alpha = \tan^{-1}(e^{\frac{2\sin(\omega_1{t})}{\omega_1}} \tan{\alpha(0)})$, shown in Fig. \ref{fig:fig-phase}B as periodic orbits.
Meanwhile, for $\omega_1 = 0$, $\omega_2 \neq 0$, elliptical periodic orbits form as shown in Fig.~\ref{fig:fig-phase}C. Here, the periods of $I(t)$, $O(t)$, and $P(t)$, denoted by $T_I$, $T_O$, and $T_P$, respectively, satisfy $T_I = T_O = T_P$, with phase shifts confirmed between them.
An attractor
form when $\omega_2 \leq 1$,
corresponding to Adler-type equation ($\omega_1=0$).
Furthermore, when $\omega_2/\omega_1 = 1/2$, limit cycles form in the $O$-$P$ plane, while $I$ remains at 1 over long time intervals (Fig.~\ref{fig:fig-phase}D). This represents special behavior within the band region of Fig.~\ref{fig:fig-2}.
Even in this band where complete aggregation is achieved ($I=1$), the discrepancy between the external field and the (completely aggregated) phase elements continues to change over time, showing that the prediction of the external field from phase elements is challenging.
Finally, for $\omega_1 = 3\pi$, $\omega_2 = 7\pi$ (Fig.~\ref{fig:fig-phase}E) and $\omega_1 = 8\pi$, $\omega_2 = 2\pi$ (Fig.~\ref{fig:fig-phase}F), respective trajectories are composed by superposition of basic periodic orbits seen in Figs. \ref{fig:fig-phase}C and \ref{fig:fig-phase}B. Such trajectories have periodicity but are structurally more complex, exhibiting apparently quasi-periodic behavior. Except for the parameter regions showing typical trajectories in Figs.~\ref{fig:fig-phase}A–D, more complex trajectories were confirmed to form.

\subsection{Extension to Rotational flow}
Having considered the aggregation dynamics for irrotational flows in the previous sections, we now extend our analysis to rotational flows that include vorticity. For this purpose, consider the coordinate transformation between an inertial frame $\mathcal{I}$ and a rotating frame $\mathcal{R}$ with angular velocity $\Omega(t)$. Let $\phi_{\mathcal{I}}(t)$ denote the phase of the external attractive point in the inertial frame. In the rotating frame, the effective phase becomes
\[
\phi_{\mathcal{R}}(t) = \phi_{\mathcal{I}}(t) - \int_0^t \Omega(s)\, ds.
\]

If $\Omega(t)$ is constant $\Omega_0$ and $\phi_{\mathcal{I}}(t) = \omega_2^{\mathcal{I}} t$, then the apparent angular velocity in the rotating frame becomes $\omega_2^{\mathcal{R}} = \omega_2^{\mathcal{I}} - \Omega_0$.
This indicates that the dynamical structure of the system is invariant under such coordinate transformations, and any resulting changes correspond to a parallel translation of the entire pattern along the $\omega_2$-axis in parameter space (Fig.~\ref{fig:fig-2}).

To demonstrate the physical implication, consider a steady irrotational flow in the inertial frame with constant stretching: $\lambda(t) = 1$ and $\phi_{\mathcal{I}}(t) = 0$, i.e., $(\omega_1^{\mathcal{I}},\omega_2^{\mathcal{I}}) = (0,0)$.
Suppose now that a uniform vorticity tensor, defined by $\Omega_{ij} = \tfrac{1}{2}(\partial_j u_i - \partial_i u_j)$, is added to the steady irrotational flow. $\Omega_{ij}$ have antisymmetric with nonzero components $(\Omega_{xy},\Omega_{yx})=(-\hat{\Omega}(t),\hat{\Omega}(t))$.We consider the case $\hat{\Omega}(t)=-1$, which corresponds to a planar Couette flow aligned parallel to the $y = -x$ direction.
This flow configuration, while involving a background $\Omega_{ij}$, is dynamically equivalent to a point at $(\omega_1^{\mathcal{R}}, \omega_2^{\mathcal{R}}) = (0, 1)$ in the parameter diagram defined in the rotating frame.
For comparison, consider vortex filaments in the inertial frame, which can be represented by the parameter point $(\omega_1^{\mathcal{I}},\omega_2^{\mathcal{I}})=(0,1)$.
From the flake dynamics perspective, the vortex filaments are indistinguishable from the Couette flow discussed above, demonstrating the equivalence of different physical flows under coordinate transformations.

We will further extend Eq.(1) along the above discussion. 
Suppose a complete aggregation state is observed in the irrotational flow under parameters $(\omega_1^{\mathcal{I}}, \omega_2^{\mathcal{I}}) = (\tilde{\omega}_1, \tilde{\omega}_2)$. Then, even when a time-dependent rotational component $\hat{\Omega}(t)$ is added to the system, the same dynamical behavior can be preserved by
reframing an angular velocity in the inertial frame as
\begin{equation*}
\omega_2^{\mathcal{I}}(t) = \tilde{\omega}_2 + \hat{\Omega}(t),
\end{equation*}
which ensures that the effective angular velocity in the rotating frame remains
\begin{equation*}
\omega_2^{\mathcal{R}}(t) = \omega_2^{\mathcal{I}}(t) - \hat{\Omega}(t) = \tilde{\omega}_2.
\end{equation*}
Thus, the model dynamics are fundamentally determined by the relative phase difference $\alpha - \phi(t)$, and remain structurally invariant under rotational transformations.
This means that for any rotational flow with $\hat{\Omega}(t)$, the evolution of $\alpha$ is extended to $\dot{\alpha}=\hat{\Omega}(t)+\lambda(t)\sin{2(\alpha-\phi(t))}$.

\section{Conclusion}

We performed theoretical and numerical analyses of phase element motion on the unit circle driven by external fields.
Motivated by flow visualization experiments using reflective flakes suspended in water waves, where flake orientations reveal local flow structures, we aimed to connect the dynamics of phase elements with observable anisotropic patterns.
The targeted external fields have structures where attractive points rotate and switch temporally, possessing two corresponding independent periods, corresponding to $\omega_1$ and $\omega_2$. It was confirmed that, in specific parameter regimes, the distribution of phases converges to a complete aggregated and stable state.

We explored complete aggregation states in multi-element systems.
We confirmed that the complete aggregation is realized in the contiguous wedge-shaped regions with a band with slope 1/2 on the $(\omega_1,\omega_2)$ plane, which is related to Arnold tongues.
This pattern reflects underlying resonance and locking phenomena, emerging under specific frequency ratios rather than as numerical artifacts.
Although aggregation and synchronization represent different dynamical concepts, our study of flake orientation in two-dimensional flows demonstrates their underlying relationship, bridging these distinct theoretical frameworks.

We introduced the reduced space $\mathbb{A}$ which presents a framework for interpreting dynamical histories and external field relationships from observable statistics. Trajectories in $\mathbb{A}$ reflect history-dependent responses, allowing inference of collective dynamics even from limited measurements, as in flake-based flow visualization. This framework is applicable to real environments where measurement is difficult and to socio-economic systems, showing that space spanned by $(I,O,P)$ may function as an effective description space for information processing beyond physical systems.

The framework extends naturally to rotational flows through the generalized equation, where vorticity effects appear as coordinate-dependent shifts, enabling application to broader classes of fluid systems beyond irrotational flows.

Future work should focus on experimental validation using controlled wave tank experiments to test the predicted critical frequency ratios.
Extension to three-dimensional flows and inclusion of inter-particle interactions would bridge our model toward more realistic flake dynamics, potentially revealing richer aggregation phenomena beyond the current two-dimensional flow assumption.

\begin{acknowledgment}
This work was supported in part by a Grant-in-Aid for Scientific Research(C) and JSPS KAKENHI Grant No.24K07331 and by the Kansai University Grant-in-Aid for progress of research in graduate course, 2024.
The authors would like to thank Mr. Yoshikawa for his valuable comments on the draft.
They also acknowledge ORDIST in Kansai University and the RIMS Joint Research Activities in Kyoto University for providing a space for their research and communication.
\end{acknowledgment}

\bibliographystyle{ieeetran}
\bibliography{flake25}

\end{document}